\documentclass[aps,pre,twocolumn,superscriptaddress,amsmath,amssymb]{revtex4}

\usepackage{graphicx}% Include figure files
\usepackage{dcolumn}% Align table columns on decimal point
\usepackage{bm}% bold math
\usepackage{color}
\usepackage{multirow}
\usepackage{textcomp}
\usepackage{umoline}
\begin{document}
\title{A greedy-navigator approach to navigable city plans}
%\author{Sang Hoon Lee\inst{1}\fnmsep\thanks{\email{sanghoon.lee@physics.umu.se}} \and Petter Holme\inst{1,2,3}}
%
%\institute{IceLab, Department of Physics, Ume{\aa} University, 901 87 Ume{\aa}, Sweden \and Department of Energy Science, Sungkyunkwan University, Suwon 440--746, Korea \and Department of Sociology, Stockholm University, 106 91 Stockholm, Sweden}
%
\author{Sang Hoon Lee}
\email{lee@maths.ox.ac.uk}
\affiliation{IceLab, Department of Physics, Ume{\aa} University, 901 87 Ume{\aa}, Sweden}
\affiliation{Oxford Centre for Industrial and Applied Mathematics, Mathematical Institute, University of Oxford, Oxford OX1 3LB, United Kingdom}

\author{Petter Holme}
\affiliation{IceLab, Department of Physics,
Ume{\aa} University, 901 87 Ume{\aa}, Sweden}
\affiliation{Department of Energy Science, Sungkyunkwan University, Suwon 440--746, Korea}
\affiliation{Department of Sociology, Stockholm University, 106 91 Stockholm, Sweden}

%\abstract{
\begin{abstract}
We use a set of four theoretical navigability indices for street maps to investigate the shape of the resulting street networks, if they are grown by optimizing these indices. The indices compare the performance of simulated navigators (having a partial information about the surroundings, like humans in many real situations) to the performance of optimally navigating individuals. We  show that our simple greedy shortcut construction strategy generates the emerging structures that are different from real road network, but not inconceivable. The resulting city plans, for all navigation indices, share common qualitative properties such as the tendency for triangular blocks to appear, while the more quantitative features, such as degree distributions and clustering, are characteristically different depending on the type of metrics and routing strategies. We show that it is the type of metrics used which determines the overall shapes characterized by structural heterogeneity, but the routing schemes contribute to more subtle details of locality, which is more emphasized in case of unrestricted connections when the edge crossing is allowed.
\end{abstract}
%} %end of abstract
%
\maketitle
\section{Introduction}
\label{intro}
Making cities easy to navigate without maps or electronic devices is a desirable objective (although certainly not the only one)
for urban planning. Based on models of how humans find their way in a partially unknown environment, one can evaluate
street maps and choose where to put new streets to optimize them for better navigability. The setup for such a simulation is
to assume every pair of start and finish points for a navigator with partial information and measure the ratio between
the shortest path length and the actual path length. In the previous work of ours, 
an index defined by the ratio is within the range $(0,1]$ with $0$ representing
a worst and $1$ a perfect navigability of {\em given} spatial graph layouts~\cite{GSN}. In this work, we instead focus on
{\em designing} optimal transport systems for such concept of navigability
under limited resources, which is an important engineering problem
and, mathematically, a typical example of the constraint optimization problem~\cite{HartmannBook}
where it is well-known that finding the exact optimum of such systems is hard.
The problem gets even more complicated if the measure or object function to be
optimized is not exactly given. Constructing transport networks on which real navigators move precisely
correspond to this situation, due to the fact that the navigability considering real navigators' behavior
is not given as a simple mathematical description~\cite{GSN,wolbers,Thomas2007,Kleinberg2000,Boguna2008,Boguna2009,Moussaid2011}.

Our idea of modeling real navigators' behavior is essentially exploiting
the spatial information on a local level, and thus it can be modeled as a simple greedy routing strategy~\cite{GSN,Kleinberg2000,Boguna2008}.
In this respect, our main idea on this work is to use the performance of a simple greedy routing scheme called
greedy spatial navigation (GSN) based on the directional information that we have introduced in Ref.~\cite{GSN},
along with the real shortest path navigation (SPN) for comparison,
to optimize the interconnected structures for a given distribution of vertices on
two-dimensional (2D) space under limited resources, i.e., the total length of edges~\cite{Gastner2006,GLi2010,Brede2010,YHu2011,WLiu2011}.
In GSN, briefly speaking, agents move in a direction as close
as possible to the direction of the target. If they reach a dead end, they backtrack to the previous point where they
have an untested choice of route. The agents in SPN, on the other hand, always follow the optimal route minimizing the
total path distance connecting the source and the target. 
Instead of solving the mathematically cumbersome nondeterministic polynomial time (NP)-complete problem
of finding the exactly optimal configurations, we focus on the realistic approach of constructing
{\em shortcuts} greedily from a spanning tree structure as the skeleton~\cite{KIGoh2006}.

Based on the simulation results from a simple shortcut construction scheme
starting with randomly distributed vertices to be connected, we first find that
the type of metric (hopping vs.~Euclidean distance) is crucial to determine
the final structures. In addition, the type of navigability (shortest vs.~GSN path length)
also plays an important role, which is reflected by the different degree of inefficiency
caused by greedy navigators between those cases. The difference between the
two navigability measures is quite prominent if we allow the crossing among
edges, which leads to the unrealistic edge condensation for the shortest
path length but only the fat-tailed degree distribution for the GSN path length.

\section{Shortcut construction}
\label{GSC_model}

The aim of the model is to construct navigability-friendly structures
under the constraint of resources given by the total length of the edges.
This fully deterministic model only depends on the initial configuration of
vertices without any free parameter involved.
First, assume that we have a set of vertices in a 2D space with their coordinates given,
without any prescribed edges.
To guarantee the connectivity and minimum initial resource, the minimum spanning tree (MST) $T$,
which is the connected subtree minimizing the total length of edges, is generated with
Kruskal's algorithm~\cite{MST}. This $T$ is the initial state of
the evolving graph $G$ so that $G(t=0) = T$.
We define the navigability as the path length averaged over all the pairs of
vertices as sources and targets.
 
The navigability is first classified as the one using
GSN pathways and the other using the real shortest path (global optimum). 
Suppose that the network is represented as a graph of $N$ vertices at coordinates $\mathbf{r}_1,\dots,\mathbf{r}_N=(x_1,y_1),\dots ,(x_N,y_N)$ that are connected by $M$ edges. Assume an agent stands at a vertex $i$ and wants to travel to $t$. Let $\mathbf{v}_{i,j}=\mathbf{r}_j-\mathbf{r}_i$ be the vector between vertices $i$ and $j$ and $\theta_j$ be the angle between $\mathbf{v}_{i,t}$ and $\mathbf{v}_{i,j}$. In case of SPN, an agent simply follows the shortest path between $i$ and $t$ using the entire geometric and topological information. An agent of GSN, in contrast, moves to the neighbor $j$ of $i$ that has not been visited before and has the smallest $\theta_j$. If all the neighbors of $i$ have been visited the navigator goes back to the vertex from which the navigator arrived to $i$, which is in contrast to the simple greedy navigation based on the geometric proximity that sometimes fails to reach $t$ due to the lack of such a backtracking process~\cite{Kleinberg2000,Boguna2008}. This procedure is repeated until $t$ is reached. 

Furthermore, we here distinguish the metrics based on the hopping distance
(the number of hops needed to reach the target) and the Euclidean distance (the
sum of Euclidean distance along the path). Note that the term ``Euclidean distance''
usually refers to the distance between two points in a metric space regardless of their
direct connection, but in this work we use this terminology as the sum of
Euclidean distance along the connected pathways, in a more flexible way.
The former is appropriate for the situation
when the time spent on the pathway is relatively insensitive to the length of each
segment of pathways, or the waiting time for each junction (vertex in this case)
is significant, e.g., the airline network or express way network with significant delays
in the junctions due to the traffic light. On the other hand, the latter is
more suitable for the case when the junction does not play a significant role, e.g.,
the road network without severe traffic and traffic light.
Such a distinction between vehicular and pedestrian patterns is responsible
for important structural difference that we will discuss in details later.
Distinction between GSN and SPN representing whether navigators only use the directional
information or have the ability to fully access the real shortest path, combined with
those two different metrics, yields four different navigation strategies in total.

\begin{figure*}
\resizebox{0.80\textwidth}{!}{
\includegraphics{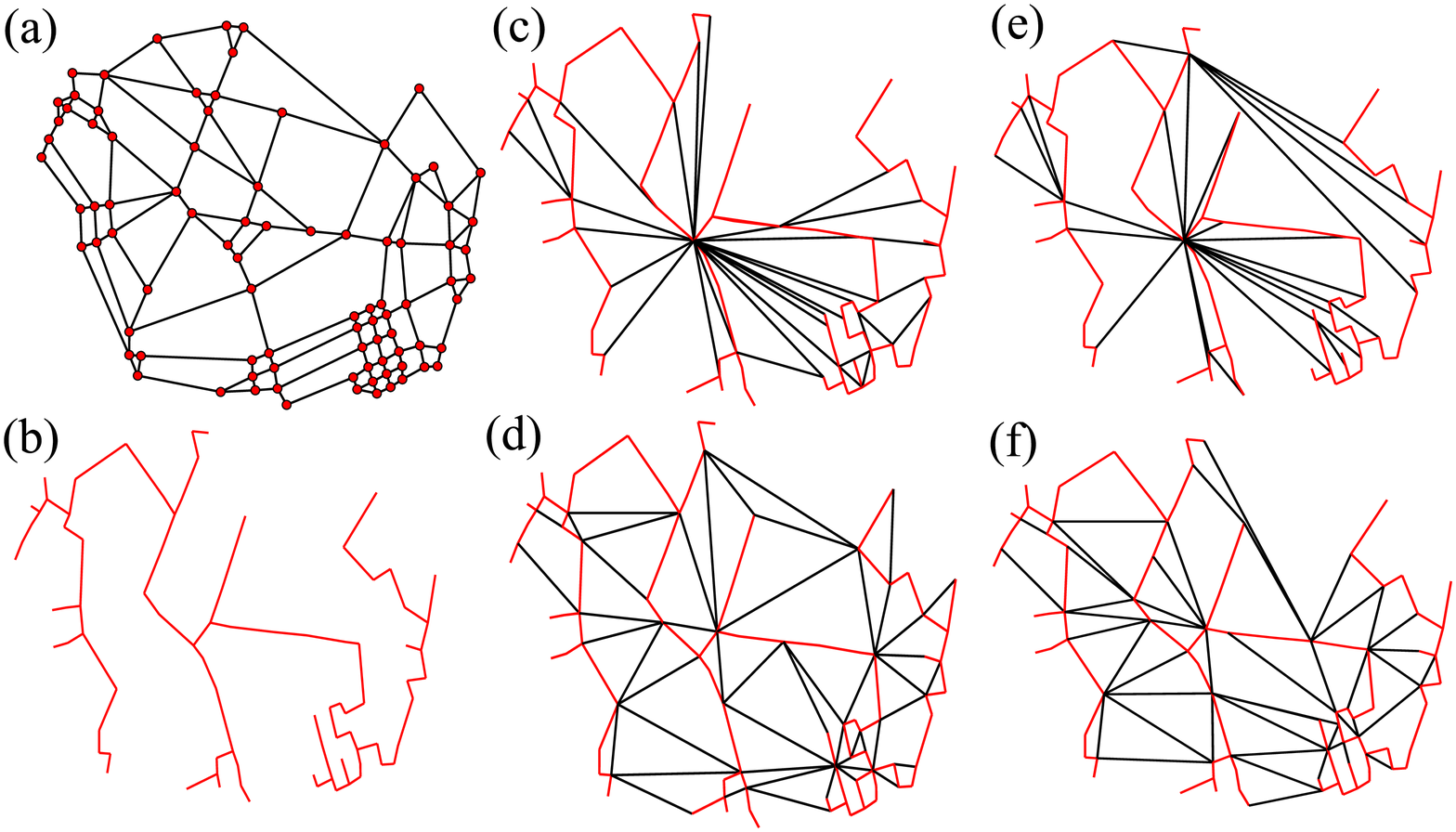}}
\caption{Original Boston road structure (a),
MST (b), GSNH (c), GSNE (d), SPNH (e), and SPNE (f). Red edges
are the ones in MST (b), and black edges correspond
to shortcut edges added during the evolving process, for (c)-(f).
Note that the edges of MST (b) in this case are chosen among
the original edges (a).
}
\label{boston_example}
\end{figure*}

At each time step $t$, the ``shortcut'' among the vertex pair (not already
connected by an edge in $G(t)$), which maximizes the performance of one of
the four navigation strategies: Greedy Spatial Navigation with Hopping distance (GSNH)~\cite{GSN},
Greedy Spatial Navigation with Euclidean distance (GSNE),
Shortest Path Navigation with Hopping distance (SPNH),
and Shortest Path Navigation with Euclidean distance (SPNE), is selected.
The shortcut is connected by a new edge, unless the candidate edge
crosses one of the existing edges of $G(t)$ in the 2D space~\cite{NoCrossingRule}.
This shortcut construction process is repeated as long as the total length
$l(t)$ of edges of $G(t)$ does not exceed a length constraint $l_\textrm{max}$.
Figure~\ref{boston_example} illustrates the shortcut construction model
based on the set of vertices in the major thoroughfares of Boston road structure~\cite{GSN,HYoun2007}.
The constraint $l_\textrm{max}$, in this case, is given by the total length of the
original edges in Fig.~\ref{boston_example}(a), and the performance
is summarized in Table~\ref{table1}. From Fig.~\ref{boston_example}, we
observe that the type of metric [hopping distance for (c) and (e) vs.~Euclidean distance
for (d) and (f)] greatly affects the final structure of the network.

\begin{table}
\caption{Total edge lengths and routing performances for the
Boston road example depicted in Fig.~\ref{boston_example}. The total
edge lengths and the GSN/SP distances are shown in an arbitrary unit.}
\label{table1}
\begin{tabular}{c|c|c}
\hline\hline
network & length & routing performance \\
\hline
original [Fig.~\ref{boston_example}(a)] & $1.038 \times 10^{5}$ & $6.820$ (GSN steps) \\
 & & $4.619 \times 10^{3}$ (GSN distance) \\
 & & $5.716$ (SP steps) \\
 & & $4.259 \times 10^{3}$ (SP distance) \\
\hline
MST [Fig.~\ref{boston_example}(b)] & $3.818 \times 10^{4}$ & $25.78$ (GSN steps) \\
 & & $1.413 \times 10^{4}$ (GSN distance) \\
 & & $17.02$ (SP steps) \\
 & & $8.173 \times 10^{3}$ (SP distance) \\
\hline
GSNH [Fig.~\ref{boston_example}(c)] & $1.036 \times 10^{5}$ & $5.266$ (GSN steps) \\
\hline
GSNE [Fig.~\ref{boston_example}(d)] & $1.029 \times 10^{5}$ & $4.228 \times 10^{3}$ (GSN distance) \\
\hline
SPNH [Fig.~\ref{boston_example}(e)] & $1.037 \times 10^{5}$ & $3.950$ (SP steps) \\
\hline
SPNE [Fig.~\ref{boston_example}(f)] & $1.029 \times 10^{5}$ & $4.103 \times 10^{3}$ (SP distance) \\
\hline\hline
\end{tabular}
\end{table}

\section{Results}
\label{sec:results}
\subsection{Topological properties of emerged network structures}

\begin{figure*}
\resizebox{0.50\textwidth}{!}{
\includegraphics{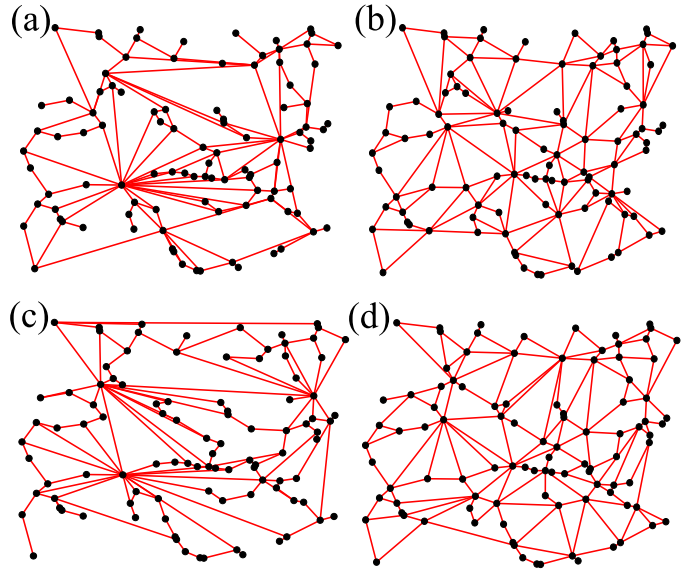}}
\caption{Example structures of model networks for
GSNH (a), GSNE (b), SPNH (c), and SPNE (d), starting from the same
randomly distributed vertices with $N = 10^2$ and $l_\textrm{max} = 20$.
}
\label{random_example}
\end{figure*}

\begin{figure*}
\includegraphics[width=\textwidth]{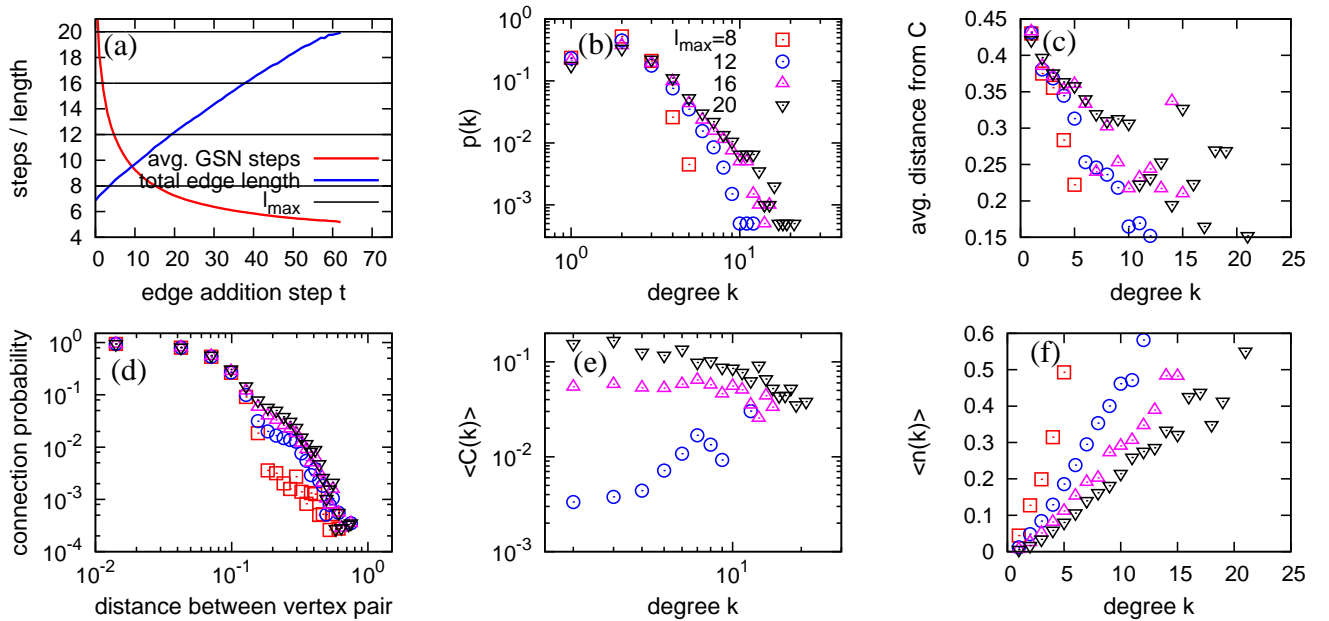}
\caption{The topological properties of emerged network structures for several different cutoff values of $l_\textrm{max}$,
in case of GSNH, where $N = 10^2$. Time series of performance and total edge length
is shown (a), where the black horizontal lines correspond to $l_\textrm{max}$ values used
in (b)--(f). The others are degree distribution (b), position centrality of vertices as
functions of degree (c), connection probability of vertex pairs as functions
of Euclidean distance (d), average clustering coefficient for vertices with given
degree (e), and average vertex navigator centrality for vertices with given degree (f),
depending on the change of $l_\textrm{max}$ values.}
\label{N100_GSN_hop_lmax_change}
\end{figure*}

\begin{figure*}
\includegraphics[width=\textwidth]{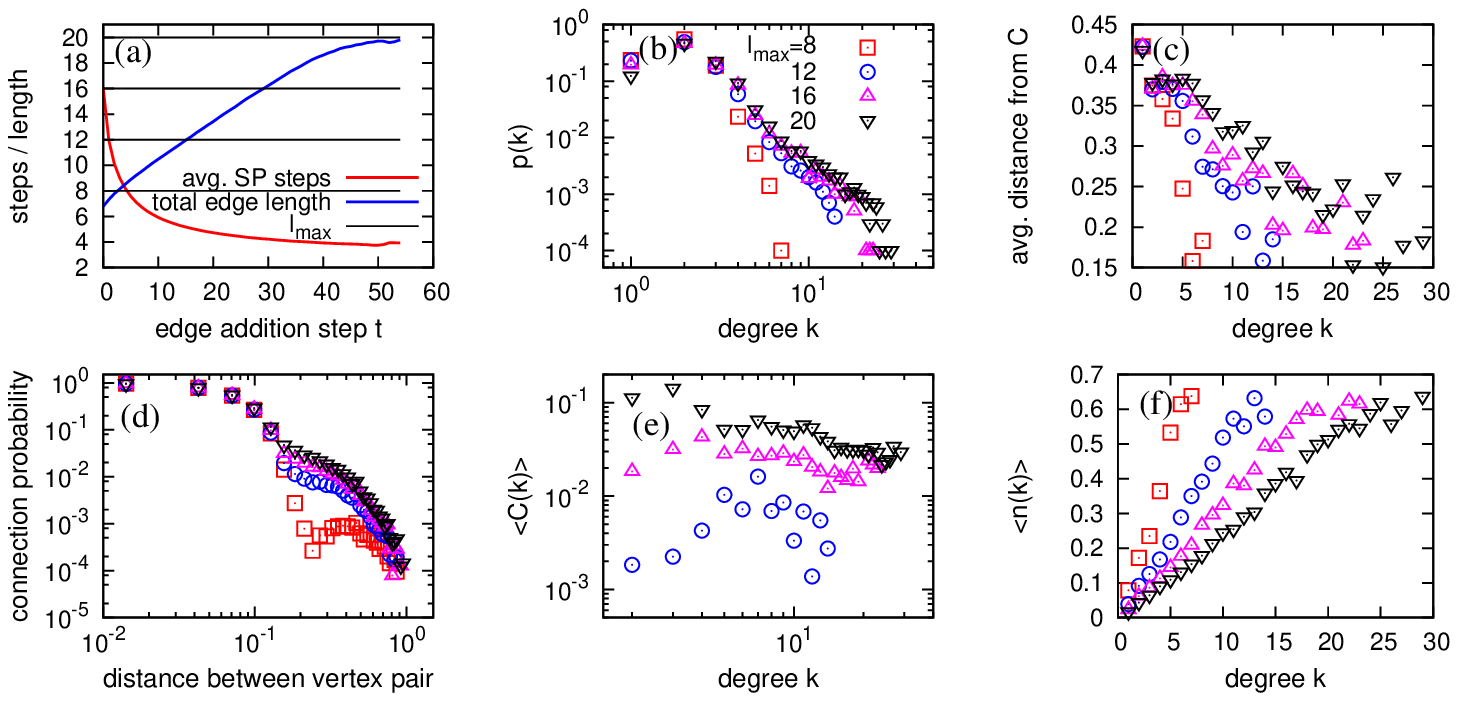}
\caption{The topological properties of emerged network structures for several different cutoff values of $l_\textrm{max}$,
as in Fig.~\ref{N100_GSN_hop_lmax_change}, in case of SPNH.
}
\label{N100_SP_hop_lmax_change}
\end{figure*}

\begin{figure*}
\includegraphics[width=\textwidth]{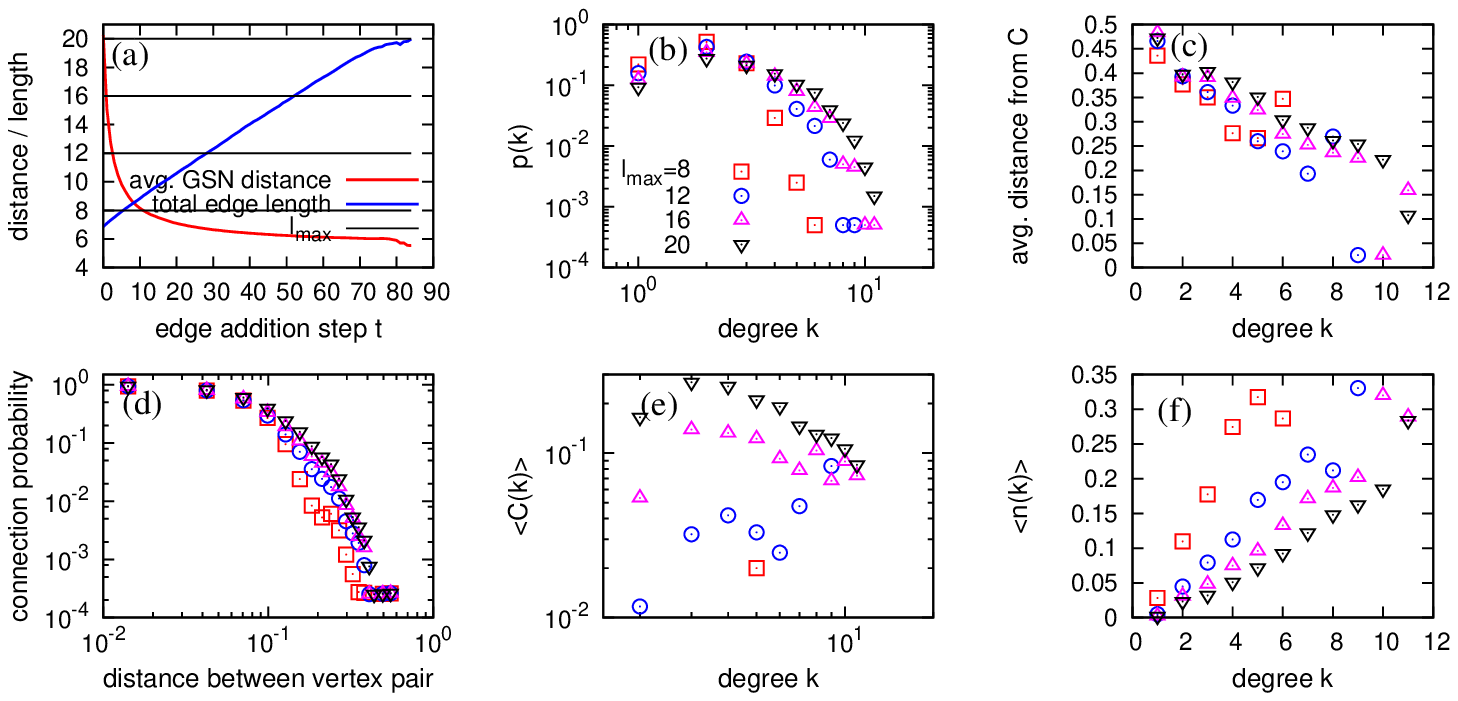}
\caption{The topological properties of emerged network structures for several different cutoff values of $l_\textrm{max}$,
as in Fig.~\ref{N100_GSN_hop_lmax_change}, in case of GSNE.
}
\label{N100_GSN_Euclidean_lmax_change}
\end{figure*}

\begin{figure*}
\includegraphics[width=\textwidth]{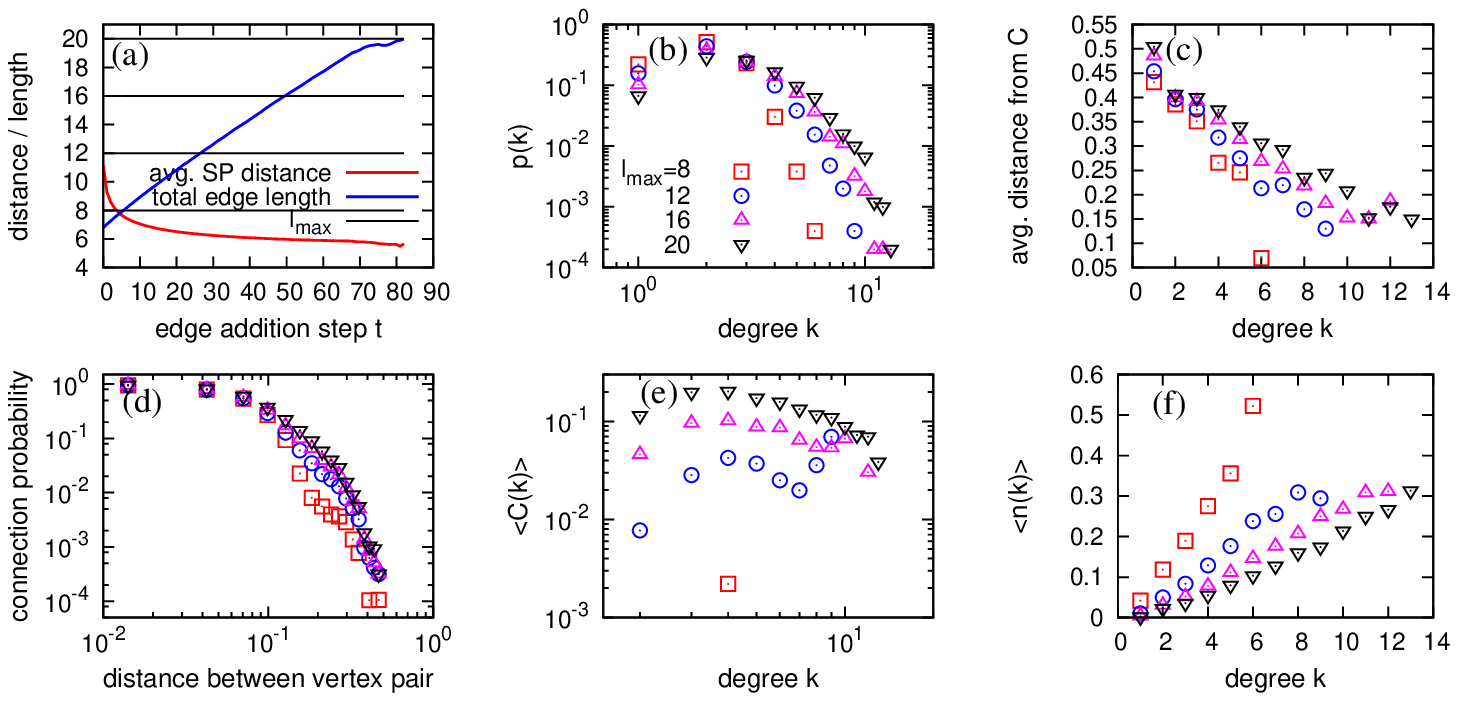}
\caption{The topological properties of emerged network structures for several different cutoff values of $l_\textrm{max}$,
as in Fig.~\ref{N100_GSN_hop_lmax_change}, in case of SPNE.
}
\label{N100_SP_Euclidean_lmax_change}
\end{figure*}

For more systematic approach, we first generate $N$ number of vertices
on the square with the unit length. Then, as described in Sect.~\ref{GSC_model},
the MST is constructed and serves as a starting point of the shortcut construction.
Examples of constructed networks from a vertex configuration in the space
are illustrated in Fig.~\ref{random_example}.
All the results reported here are from graphs with $N = 10^2$ and averaged at least $20$ independent graph generations.
At the first glance, the difference between GSN and SP [(a) vs.~(c) and (b) vs.~(d)] may not be notable enough,
but the fraction of Braess edges (defined as the edges whose removal enhances the greedy navigability)~\cite{GSN}
causing the inefficiency, shown in
Table~\ref{table2}, is much smaller for GSN, from its construction purpose
toward the optimized structure for the GSN pathways.
The comparative time series of performance and the total edge length shown in
Figs.~\ref{N100_GSN_hop_lmax_change}--\ref{N100_SP_Euclidean_lmax_change}(a) indicate that the GSNH tends to connect
edges with shorter distance first.

\begin{table}
\caption{Fraction of Braess edges $P_B$ with negative edge essentiality $e$ values~\cite{GSN} for
emerged networks with different greedy shortcut addition schemes, for the model network
with $N = 10^2$ and $l_\textrm{max} = 20$. The $e$ values defined with the
hopping (Euclidean) distance are used for GSNH and SPNH (GSNE and SPNE), respectively.}
\label{table2}
\begin{tabular}{c|cccc}
\hline\hline
& GSNH & GSNE & SPNH & SPNE \\
\hline
$P_B$ & $12.12\%$ & $1.249\%$ & $21.70\%$ & $5.215\%$ \\
\hline\hline
\end{tabular}
\end{table}

The most noticeable feature distinguishing between using the hopping distance and the Euclidean distance
is the heterogeneity in the number of connections attached to each node, or degree distribution shown
in Figs.~\ref{N100_GSN_hop_lmax_change}--\ref{N100_SP_Euclidean_lmax_change}(b).
Emergence of hubs (vertices with large degrees) for hopping-distance-based scheme is 
clearly evident as shown in much fatter tail from
Figs.~\ref{N100_GSN_hop_lmax_change}(b) and \ref{N100_SP_hop_lmax_change}(b) than
Figs.~\ref{N100_GSN_Euclidean_lmax_change}(b) and \ref{N100_SP_Euclidean_lmax_change}(b).
%(``hub emergence based on particle dynamics'' in Ref.~\cite{SWKim2008}),
Looking more closely, it is also shown that even some large hubs located relatively
far from the centroid, showing the central governance (SPN) vs.~decentralized local governance (GSN)
hub locality [Figs.~\ref{N100_GSN_hop_lmax_change}--\ref{N100_SP_Euclidean_lmax_change}(c)].
The different importance of geometric distance for routing schemes is reflected in
the average connection probability for vertex pairs with
a certain Euclidean distance, shown in Figs.~\ref{N100_GSN_hop_lmax_change}--\ref{N100_SP_Euclidean_lmax_change}(d),
where the connection probability is exponentially decreased for GSNE and SPNE, while the relatively long-range connections are
observed for GSNH and SPNH.
The average clustering coefficient $\langle C(k) \rangle$~\cite{Watts1998} as a function of degrees seem to be increased with
the degree [Figs.~\ref{N100_GSN_hop_lmax_change}--\ref{N100_SP_Euclidean_lmax_change}(e)], in contrast to
topological networks without geometric embedding.
In addition, naturally, there is a strong correlation between the vertex navigator centrality $\langle n(k) \rangle$~\cite{GSN} and degrees [Figs.~\ref{N100_GSN_hop_lmax_change}--\ref{N100_SP_Euclidean_lmax_change}(f)].

\subsection{Geometric properties: triangular block, angle and area distributions}

One can observe lots of characteristic triangular blocks for the optimized
structures for all the cases, as shown in Figs.~\ref{boston_example} and \ref{random_example}.
Those triangular blocks are quantitatively counted in comparison to the random
counterparts, using the clustering coefficient of a whole network~\cite{CC_triangle}.
Table~\ref{table3} shows the results. Based on the ratio $C_{\bigtriangleup}/C_{r}$
where $C_{\bigtriangleup}$ is the actual number of triangles and $C_{r} = 2M/N^{2}$
is the expected number of triangles if all the triangles are formed just by
chance,
obviously the Euclidean-distance based strategy produces more triangular blocks,
but for both metrics, it is also notable that GSN induces more triangles than
SPN. Therefore, along with the degree distributions, the triangular block statistics
also support the conclusion that GSN encapsulates the refined local structures.
In reality, of course, due to other factors such as packing the buildings with
square cross-sections into blocks, the square blocks rather than triangular blocks
prevail. However, it is important to note that such a simple objective function
based on navigability can generate the local block structures. Those triangular
structures are indeed observed in reality, such as historical centers of cities built
by a pedestrian strategy~\cite{Buhl2006,Strano2012}.
Such a difference is shown to be even more prominent when we remove the no-crossing
rule, which will be discussed in Sect.~\ref{no_crossing_rule}.

\begin{table}
\caption{The clustering coefficient based on the number of triangles ($C_{\bigtriangleup}$),
compared to the random counterpart ($C_{r} = 2M/N^2$, where $N$ and $M$ are
the numbers of vertices and edges, respectively).}
\label{table3}
\begin{tabular}{c|c|c|c}
\hline\hline
method & $C_{\bigtriangleup}$ & $C_{r}$ & $C_{\bigtriangleup}/C_{r}$ \\
\hline
GSNH & $1.04 \times 10^{-1}$ & $3.20 \times 10^{-2}$ & $3.26$ \\
\hline
GSNE & $1.89 \times 10^{-1}$ & $3.66 \times 10^{-2}$ & $5.15$ \\
\hline
SPNH & $6.29 \times 10^{-2}$ & $2.98 \times 10^{-2}$ & $2.11$ \\
\hline
SPNE & $1.56 \times 10^{-1}$ & $3.44 \times 10^{-2}$ & $4.53$ \\
\hline\hline
\end{tabular}
\end{table}

\begin{figure*}
\includegraphics[width=0.8\textwidth]{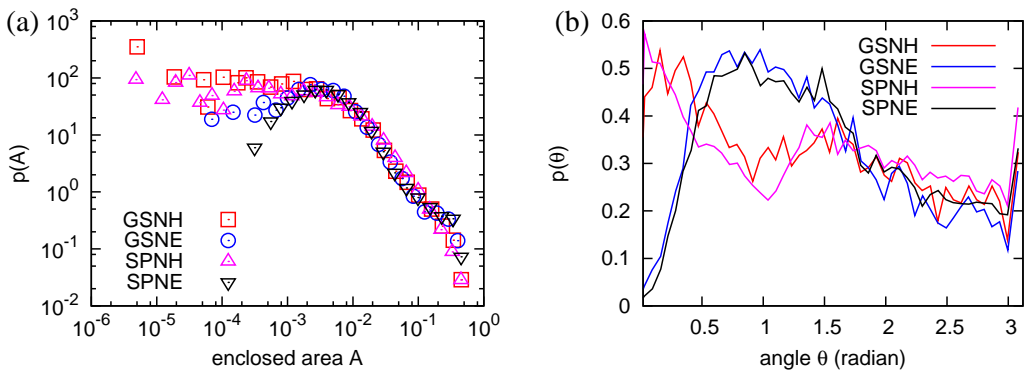}
\caption{The distributions of area enclosed by edges (a)
and the angles for the intersections (vertices) (b),
in the optimized structure for each navigability ($N = 10^2$ and
$l_{\textrm{max}} = 20$).
}
\label{enclosed_area_angle_dist}
\end{figure*}

Another geometric aspect of the optimized networks is seen by observing
the distributions of area enclosed by edges and the angles by adjacent
edges for vertices, i.e., the angles between adjacent roads
at the intersections. We observe that both the enclosed area distribution
and angle distribution are notably distinguishable depending
on the metrics used as shown in Fig.~\ref{enclosed_area_angle_dist}.
For hopping-distance based strategies, more heterogeneous enclosed area
distributions compared to the Euclidean-distance based strategies are observed (Fig.~\ref{enclosed_area_angle_dist}(a)).
In addition,
very sharp angles (Fig.~\ref{enclosed_area_angle_dist}(b)) are abundant due to the existence of hubs
(see Fig.~\ref{random_example} as well for examples), and
the tendency is slightly more significant in SPNH than GSNH. On the other
hand, the angles are distributed around $\simeq 60^{\circ}$, the characteristic
angle of the regular triangle for Euclidean-distance based GSNE and SPNE. Therefore, the enclosed area and angle distributions also
indicate the emergence of relatively more regular triangular structures with uniform enclosed area for GSNE and SPNE.

\subsection{Remarks on the no-crossing rule for edges}
\label{no_crossing_rule}

\begin{figure*}
\includegraphics[width=0.50\textwidth]{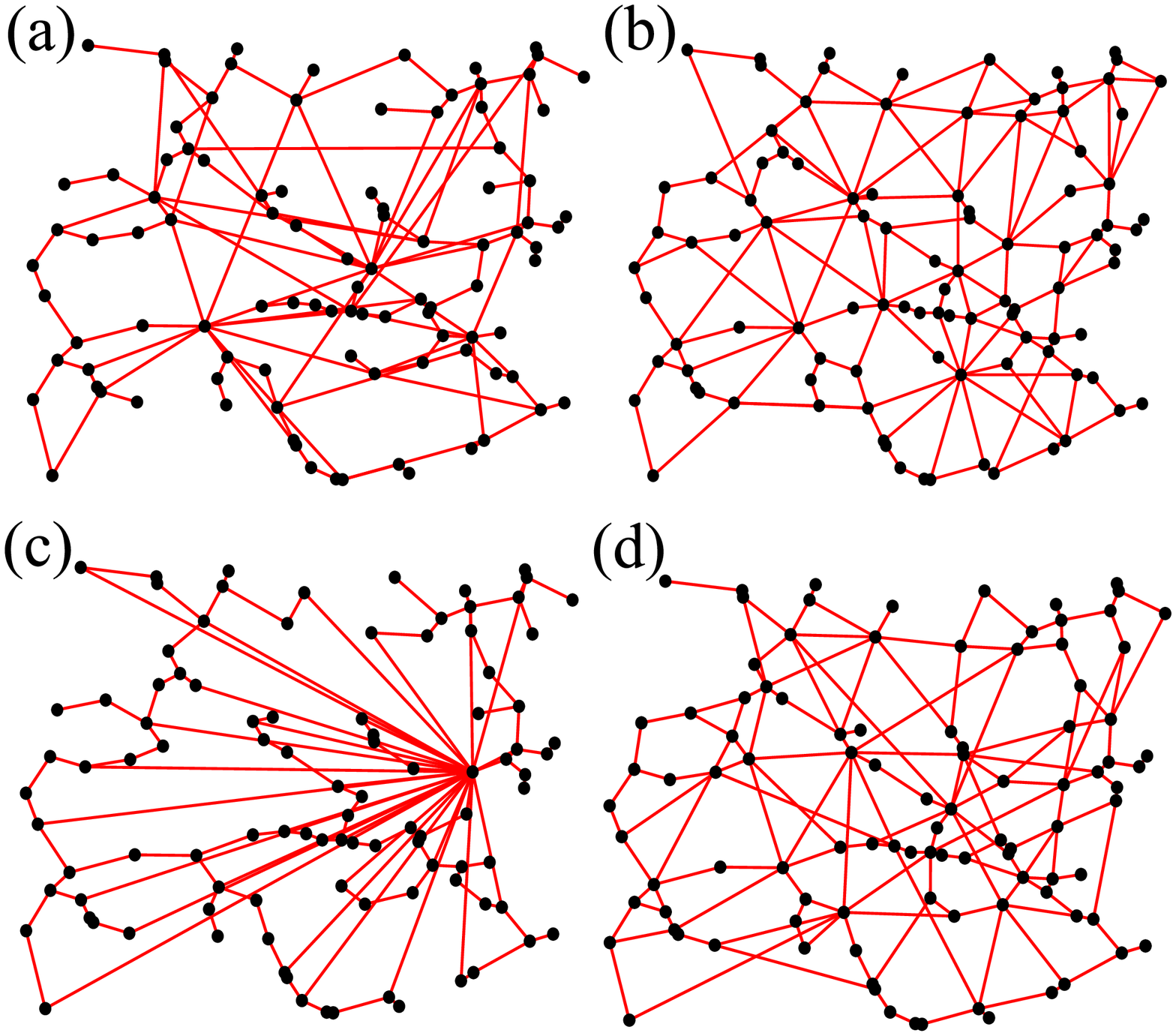}\caption{Example structures of model networks for
GSNH (a), GSNE (b), SPNH (c), and SPNE (d), if edge-crossing is allowed,
starting from the same
randomly distributed $10^2$ vertices ($l_\textrm{max} = 20$)
as in Fig.~\ref{random_example}.
}
\label{x_random_example}
\end{figure*}

\begin{figure*}
\includegraphics[width=0.8\textwidth]{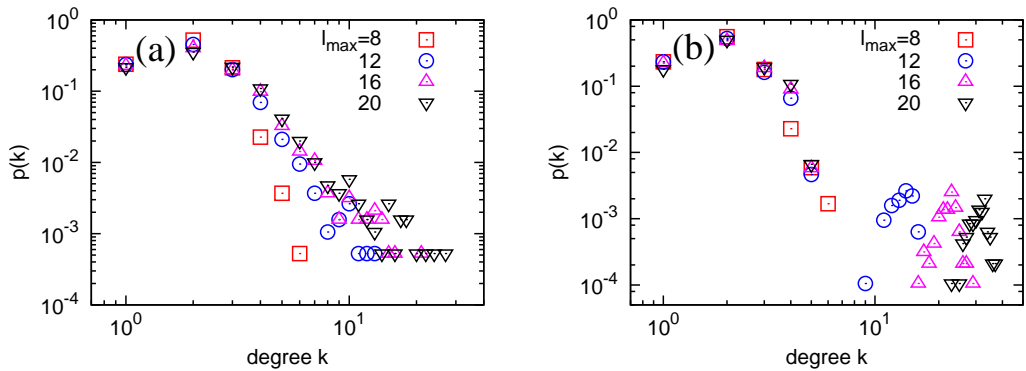}
\caption{The degree distributions of emerged network structures for GSNH (a) and SPNH (b),
for several different cutoff values of $l_\textrm{max}$, where $N = 10^2$.
}
\label{x_pk_GSNH_SPNH}
\end{figure*}

So far, we have not allowed the crossing between edges in the construction process
since we consider such a crossing as effectively generating a new junction or vertex.
What if there is no such rule?
If we allow edge-crossing, star-graph-like structures are naturally
emerged for SPNH as shown in Fig.~\ref{x_random_example}(c), which leads to the
``condensation'' in terms of degree, as the bimodal distribution for $l_{\textrm{max}} = 12, 16$, and $20$
shown in Fig.~\ref{x_pk_GSNH_SPNH}(b)~\cite{Bianconi2001}.
Interestingly, this severe condensation is not observed in case of GSNH, as shown in Figs.~\ref{x_random_example}(a)
and \ref{x_pk_GSNH_SPNH}(a) (power-law-like fat tailed distribution but no condensation).
The no-crossing rule, therefore, effectively prevents such a condensation for SPNH, while
no such explicit rule is necessary for GSNH, because the consideration of greedy navigators itself
naturally avoid such condensed situations and allows the {\em local} hubs. The combinatorial effects
of taking both hopping and Euclidean distances are discussed in Ref.~\cite{Fabrikant2002} as well.

Finally, we remark that our simple model does not take into account
the realistic number of connected roads for junctions (degree in the 
notion of graphs), which is between $2.1$ and $2.9$~\cite{Buhl2006}.
Considering such restrictions may make the model more realistic,
but we would like to emphasize that our model can be very valid for much
wider contexts, besides the street networks, without those explicit constraints.

\section{Summary and discussions}
\label{summary_discussion}
With a simple greedy optimization under the limited sum of length, we have
investigated the various structural properties of optimized networks for
different types of navigability. Due to the complexity of finding the exact
optimal configurations of transportation systems, in practice, heuristic methods
are usually adopted for real systems. Therefore, we believe that even though
our scheme is simple and idealistic, it has some degree of implications in
real transportation systems. The most essential findings of ours are the emergence
of hubs in case of the consideration of the hopping distance, inherently geometric
aspect of optimized structures reflecting the types of metrics used,
and the fact that the user-based routing scheme (GSN) effectively induces
the local hubs even without the explicit no-crossing rule. The emergence of
triangular blocks from the simple optimization procedure is also notable---would city plans be optimized for pedestrian navigability rather than traffic planning and building design, then we should probably see more triangular blocks.
Also, we would like to note that the triangular like pattern has always been found in the historical centers of cities built
by a pedestrian strategy~\cite{Buhl2006,Strano2012}.

More elaborated or realistic approaches, for instance, using the different types of backbones,
stochastic edge additions (similar to the simulated annealing process);
different objective functions such as the combination of navigability and
length~\cite{Gastner2006} or among different metrics, etc.~are good candidates for the future work.
The reality check using human subjects, 
of course, would be necessary for the application to the real systems.

%\section{Section title}
%\label{sec:1}
%and \cite{RefJ}
%\subsection{Subsection title}
%\label{sec:2}
%as required. Don't forget to give each section
%and subsection a unique label (see Sect.~\ref{sec:1}).
%

%\begin{figure}
%\resizebox{0.75\columnwidth}{!}{%
%  \includegraphics{fig1.pdf} }
%\caption{Please write your figure caption here.}
%\end{figure}
%\begin{table}
%\caption{Please write your table caption here.}
%\begin{tabular}{lll}
%\hline\noalign{\smallskip}
%\end{tabular}
%\end{table}
%

%\begin{acknowledgement}
\begin{acknowledgments}
This research is supported by the Swedish Research Council and the WCU program through NRF Korea funded by MEST R31--2008--000--10029--0 (PH).
%\end{acknowledgement}
\end{acknowledgments}


\begin{thebibliography}{00}
\bibitem{NoCrossingRule} This rule is applied since we consider the road-like
structure, where each vertex corresponds to the junctions of roads. Therefore,
a new crossing point generated by a new edge can be considered as a new vertex,
which effectively changes the system size, so we intentionally avoid this situation.
\bibitem{CC_triangle} Note that there are two slightly different kinds of clustering coefficients,
i.e., the one averaged over the individual vertices' clustering coefficient
which is used for $C(k)$ in this work, and the other taking the ratio of numbers
of triangles to the numbers of vertex triads used for $C_{\bigtriangleup}$.
\end{thebibliography}
\end{document}